\documentclass[conference]{IEEEtran}

\usepackage{cite}

\ifCLASSINFOpdf
\else
\fi

\usepackage{amsmath}
\usepackage[pdftex]{graphicx}
\usepackage{lineno}
\usepackage[bookmarks=false]{hyperref}
\usepackage{float}
\usepackage{graphicx}

\graphicspath{{./}}
\graphicspath{{./}}

\usepackage[table,xcdraw]{xcolor}
\usepackage{amsmath}
\usepackage{multirow}
\usepackage{flushend}
\usepackage{lipsum}
\usepackage{cleveref}
\usepackage{tikz}
\usepackage{balance}

\def\yC{\raisebox{-1.2pt}{\tikz\node[anchor=south,circle,scale=0.5,color=white, fill=yellow]{\textbf{H}};}}
\def\gC{\raisebox{-1.2pt}{\tikz\node[anchor=south,circle,scale=0.5,color=white, fill=green]{\textbf{U}};}}
\def\rC{\raisebox{-1.2pt}{\tikz\node[anchor=south,circle,scale=0.5,color=white, fill=red]{$\quad$};}}

\usepackage[tight]{units}
\usepackage[]{subcaption}
\usepackage[]{verbatim}
\usepackage{tabto}
\usepackage{acro}
\acsetup{first-style=all}

\DeclareAcronym{URLLC}{
  short = URLLC ,
  long  = Ultra-Reliable Low Latency Communication ,
  class = abbrev
}

\DeclareAcronym{AL}{
  short = AL ,
  long  = Aggregation Level ,
  class = abbrev
}

\DeclareAcronym{FDD}{
  short = FDD ,
  long  = Frequency Division Duplexing ,
  class = abbrev
}

\DeclareAcronym{TDD}{
  short = TDD ,
  long  = Time Division Duplexing ,
  class = abbrev
}

\DeclareAcronym{QAM}{
  short = QAM ,
  long  = Quadrature Amplitude Modulation ,
  class = abbrev
}

\DeclareAcronym{LLR}{
  short = LLR ,
  long  = Log-Likelihood Ratio ,
  class = abbrev
}

\DeclareAcronym{DMRS}{
  short = DMRS ,
  long  = Demodulation Reference Signal,
  class = abbrev
}

\DeclareAcronym{WI}{
  short = WI ,
  long  = Work Item,
  class = abbrev
}

\DeclareAcronym{E2E}{
  short = E2E ,
  long  = End-to-End ,
  class = abbrev
}

\DeclareAcronym{DL}{
  short = DL ,
  long  = downlink ,
  class = abbrev
}
\DeclareAcronym{SNR}{
  short = SNR ,
  long  = Signal-to-Noise Ratio,
  class = abbrev
}
\DeclareAcronym{CFI}{
  short = CFI ,
  long  = Control Format Indicator ,
  class = abbrev
}

\DeclareAcronym{UL}{
  short = UL ,
  long  = uplink ,
  class = abbrev
}

\DeclareAcronym{SR}{
  short = SR ,
  long  = Scheduling Request ,
  class = abbrev
}

\DeclareAcronym{BLER}{
  short = BLER ,
  long  = Block Error Rate ,
  class = abbrev
}

\DeclareAcronym{CRC}{
  short = CRC ,
  long  = Cyclic Redundancy Check ,
  class = abbrev
}

\DeclareAcronym{CCE}{
  short = CCE ,
  long  = Control Channel Element ,
  class = abbrev
}

\DeclareAcronym{PDCCH}{
  short = PDCCH ,
  long  = Physical Downlink Control CHannel ,
  class = abbrev
}

\DeclareAcronym{LTE}{
  short = LTE ,
  long  = Long Term Evolution ,
  class = abbrev
}

\DeclareAcronym{NGMN}{
  short = NGMN ,
  long  = Next Generation Mobile Networks ,
  class = abbrev
}

\DeclareAcronym{RNTI}{
  short = RNTI ,
  long  = Radio Network Temporary Identifier ,
  class = abbrev
}

\DeclareAcronym{3GPP}{
  short = 3GPP ,
  long  = 3rd Generation Partnership Project ,
  class = abbrev
}

\DeclareAcronym{HRLLC}{
  short = HRLLC ,
  long  = High-Reliable Low Latency Communication ,
  class = abbrev
}

\DeclareAcronym{TTI}{
  short = TTI ,
  long  = Transmission Time Interval ,
  class = abbrev
}

\DeclareAcronym{sTTI}{
  short = sTTI ,
  long  = short Transmission Time Interval ,
  class = abbrev
}

\DeclareAcronym{RTT}{
  short = RTT ,
  long  = Round Trip Time ,
  class = abbrev
}

\DeclareAcronym{LDPC}{
  short = LDPC ,
  long  = Low-Density Parity-Check ,
  class = abbrev
}

\DeclareAcronym{UE}{
  short = UE ,
  long  = User Equipment ,
  class = abbrev
}

\DeclareAcronym{DCI}{
  short = DCI ,
  long  = Downlink Control Information ,
  class = abbrev
}

\DeclareAcronym{HARQ}{
  short = HARQ ,
  long  = Hybrid Automatic Repeat reQuest ,
  class = abbrev
}

\DeclareAcronym{mMTC}{
  short = mMTC ,
  long  = massive Machine Type Communications ,
  class = abbrev
}

\DeclareAcronym{5G}{
  short = 5G ,
  long  = Fifth Generation ,
  class = abbrev
}

\DeclareAcronym{SPS}{
  short = SPS ,
  long  = Semi-Persistent Scheduling ,
  class = abbrev
}

\DeclareAcronym{PI}{
  short = PI ,
  long  = Pre-emption Indication ,
  class = abbrev
}

\DeclareAcronym{NR}{
  short = NR ,
  long  = New Radio ,
  class = abbrev
}

\DeclareAcronym{eMBB}{
  short = eMBB ,
  long  = enhanced Mobile BroadBand ,
  class = abbrev
}

\DeclareAcronym{angelsperarea}{
  short = \ensuremath{a} ,
  long  = The number of angels per unit area ,
  sort  = a ,
  class = nomencl
}
\DeclareAcronym{numofangels}{
  short = \ensuremath{N} ,
  long  = The number of angels per needle point ,
  sort  = N ,
  class = nomencl
}
\DeclareAcronym{areaofneedle}{
  short = \ensuremath{A} ,
  long  = The area of the needle point ,
  sort  = A ,
  class = nomencl
}

\begin{document}

\clubpenalty = 10000
\widowpenalty = 10000
\displaywidowpenalty = 10000

\bstctlcite{MyBSTcontrol}

	\title{URLLC Services in 5G \\ Low Latency Enhancements for LTE}

\author{\IEEEauthorblockN{Thomas Fehrenbach\IEEEauthorrefmark{1}, Rohit Datta\IEEEauthorrefmark{2}, Bari\c{s} G\"oktepe\IEEEauthorrefmark{1}, Thomas Wirth\IEEEauthorrefmark{1}, and Cornelius Hellge\IEEEauthorrefmark{1}}
	\IEEEauthorblockA{\IEEEauthorrefmark{1} Fraunhofer Heinrich Hertz Institute (HHI),  Berlin, Germany.}
    \IEEEauthorblockA{\IEEEauthorrefmark{2} Fraunhofer Institute for Integrated Circuits (IIS),  Erlangen, Germany.}
    }

	\maketitle

    \begin{abstract}
5G is envisioned to support three broad categories of services: eMBB, URLLC, and mMTC. URLLC services refer to future applications which require reliable data communications from one end to another, while fulfilling ultra-low latency constraints. In this paper, we highlight the requirements and mechanisms that are necessary for URLLC in LTE. Design challenges faced when reducing the latency in LTE are shown. The performance of short processing time and frame structure enhancements are analyzed. Our proposed DCI Duplication method to increase LTE control channel reliability is presented and evaluated. The feasibility of achieving low latency and high reliability for the IMT-2020 submission of LTE is shown. We further anticipate the opportunities and technical design challenges when evolving 3GPP's LTE and designing the new 5G NR standard to meet the requirements of novel URLLC services.

	\end{abstract}

	\IEEEpeerreviewmaketitle

	\section{Introduction}

\begin{figure}[b]
\vspace{-0.2cm}
\centering
  \includegraphics[width=1.00\linewidth]{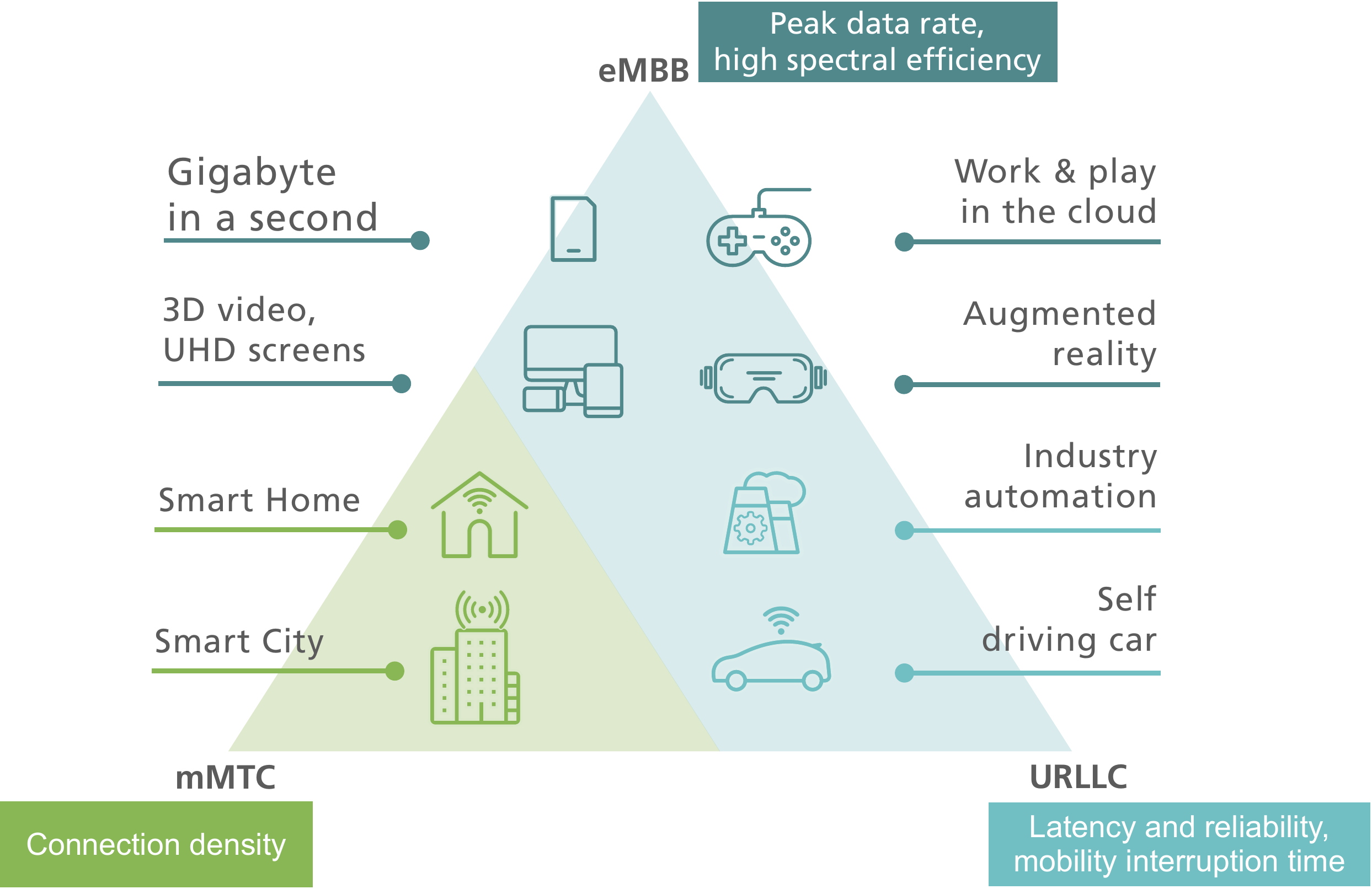}
  \caption{Use cases for IMT-2020 and beyond~\cite{itu.m.2410}.}
  \label{fig:ITUreq}
\end{figure}

The emerging 5G wireless mobile networks will be as much the result of relentless and extensive improvements of \acuse{3GPP}3GPP's~(3rd Generation Partnership Project) \ac{LTE} as it is a technology revolution~\cite{path25G}. Besides the possibilities for
self-contained subframes, an entirely new air interface or grant-free access,
it also prompts development of numerous incremental improvements.
The IMT-2020 use cases, as depicted in Fig.~\ref{fig:ITUreq}, shall fulfill three principal dimensions of performance~\cite{itu.m.2410,20155GWhitePaperNGMN}.
5G will not only focus on \ac{eMBB}; but \ac{URLLC} and \ac{mMTC} seemingly have a similar footing in long-term visions of what 5G might ultimately become~\cite{7980747}.

New use cases demanding very low latency, very high reliability or a combination of high reliability and low latency, i.e. \ac{URLLC}, have been identified as one of the key trends of future wireless cellular communications~\cite{3GP15Timeline,3gpp.RP-171489}. Such use cases include a rather diverse set of requirements on the combination of reliability and latency such as remote tactile or haptic control (low latency), wireless communications in industrial automation (high reliability, low/medium latency), and smart grids (high reliability, low/medium latency), just to mention a few.

Alongside \ac{NR}, LTE technology enhancements are needed to serve such new use cases and to remain technologically competitive up to and beyond 2020. As a candidate technology, it is motivated to further enhance the LTE system, such that the IMT-2020 5G requirements~\cite{itu.m.2410} can be met. Including those for URLLC in terms of reliability, packet loss of $10^{-5}$ for small data packets, as well as low latency of less than~\unit[1]{ms} in one way user plane.

The \ac{3GPP} LTE Rel. 14 with its \ac{WI} on \textit{L2 latency reduction} and the technical report on \textit{shortened \ac{TTI} and processing time for LTE}~\cite{3gpp.36.881} provides solutions for L1/L2 latency reduction. These solutions enable latencies at the levels mentioned above, but new functionality is needed to improve the reliability under latency constraints required for \ac{URLLC} services.
Although the term URLLC targets both achieving a very low latency, as well as fulfilling a reliability constraint, the 3GPP standardization body decoupled latency and reliability aspects. Initial focus of improving \ac{LTE} system performance was on latency related aspects and is referred to as sTTI~(short \ac{TTI}) \ac{WI} \cite{3gpp.RP-161299}. Reliability aspects were the target of a later \ac{WI} under the term \ac{HRLLC}~\cite{3gpp.RP-171489}.

An overview on standardization activities in \ac{3GPP} as an analysis of latency and simulation results on robustness for standard \ac{LTE} and using our proposed novel scheme for \ac{DCI} duplication are presented in this paper.
Section~II highlights the technical requirements for URLLC systems, followed by Section~III which describes the technical solutions developed and standardized in 3GPP in the context of \ac{LTE}.
In Section~IV we present our results and performance analysis. Latency is analyzed analytically and reliability improvements of the novel \ac{DCI} duplication approach are presented.
Section~V anticipates \ac{NR} \ac{URLLC}, and finally a conclusion is given in Section VI.

\section{Physical Layer Requirements: High vs. Ultra Reliable Low Latency}

The new generation radio system (5G) addresses the demands and business contexts of 2020 and beyond. In 2015, the \ac{NGMN} alliance published their 5G whitepaper~\cite{20155GWhitePaperNGMN}, listing the mobile operators' vision on 5G use cases, business models and requirements.

\textbf{Latency-related aspects:} The NGMN proposed that 5G systems shall be able to provide \unit[10]{ms} \ac{E2E} latency in general (referred to as HRLLC in 3GPP), and \unit[1]{ms} latency (URLLC) for use cases with extremely low latency requirements. E2E latency refers to the duration between the transmission of a small data packet from the application layer and successful reception at the application layer of the destination node. The over-the-air latency constitutes only one part of the \ac{E2E} latency, whereas the core network latency poses the residual part. Hence, 3GPP agreed on aiming for~\unit[0.5]{ms} over-the-air latency, although \unit[1]{ms} is still the hard constraint~\cite{3gpp.38.913}.

\textbf{Reliability-related aspects:} 3GPP defines the reliability by the probability to successfully transmit a packet from one radio unit to another radio unit within the given time constraint required by the targeted service \cite{3gpp.38.913}. For the sake of convenience, we describe the reliability with the complementary probability, that is the packet failure rate. For \ac{URLLC}, 3GPP defines the target packet failure rate of $10^{-5}$ within \unit[1]{ms} over-the-air latency. A more relaxed constraint of $10^{-4}$, has been defined for \ac{HRLLC}, which is a challenge for todays 4G systems. Note, 4G systems for \ac{eMBB} typically operate at a target \ac{BLER} of $10^{-1}$. Thus, future LTE releases as well as clean-slate 5G systems face tough design challenges when addressing ultra-high reliability combined with a stringent latency objective. However, the feasibility of implementing URLLC with~\unit[1]{ms} E2E latency and NR-like parameters with a Software-Defined Radio~(SDR) platform was recently shown~in~\cite{2015AdvancedSDRWirth,2016TacintPilz}.

\section{3GPP Standardization Efforts in LTE Rel. 15}

\subsection{LTE Latency Reduction Mechanisms}
Two basic mechanisms were defined in LTE Rel. 15 to reduce latency, namely reduced processing time and the support of a shortened frame structure. The latter is referred to as \acf{sTTI}\acuse{TTI}.

\textbf{Reduced processing time}: For a data packet arriving at \ac{TTI} $n$, the processing time is shortened from $n+4$ down to $n+3$.
With short processing time, the \acuse{UE}User Equipment's (UE) response time from \ac{DL} data transmission to DL \ac{HARQ} and from \ac{UL} grant to UL data transmission is reduced from $n+4$~\acp{TTI} to $n+3$~\acp{TTI}. This means that the \ac{HARQ} \ac{RTT} is reduced from $n+8$ to $n+6$ for both DL and UL.

\begin{figure}[t]
\vspace{-0.0cm}
  \includegraphics[width=0.98\linewidth]{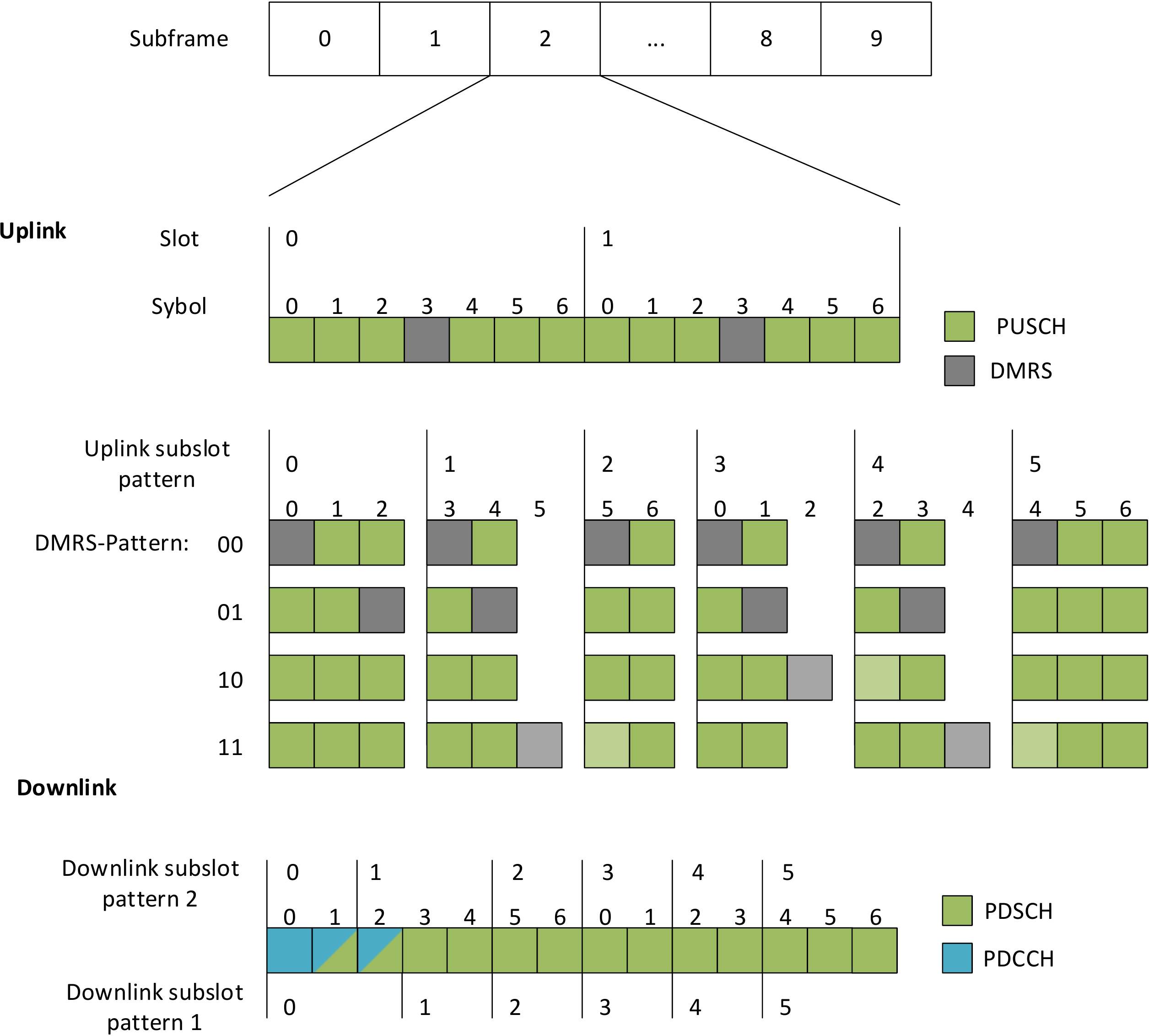}
  \caption{Frame structure type 1 (FDD) in LTE~\cite{3gpp.36.211}.}
  \label{fig:FS1}
  \vspace{-0.2cm}
\end{figure}

\textbf{Short TTI} reduces the transmission time by introducing shorter frame structure. Dividing the $1$~ms subframe into either 2 parts (slots) or 5-6 parts (subslots) as shown in Fig. \ref{fig:FS1}. For slot duration, the latencies are calculated on the assumption that the TTI is 7~symbols, whereas for subslot configuration the latencies are calculated following the subslot layout.
For slot and subslot configuration, the processing time is scaled with the TTI length. Hence, the absolute processing time is reduced by a factor of 2~for the slot, and a factor of 5-6~for the subslot configuration. Note, for slot and subslot configurations with sTTI, the processing time remains $n+4$~but scales down with the reduced TTI length.\looseness=-1
\vspace{-1mm}
\subsection{Division Duplexing and sTTI}
Introducing sTTI in LTE has conflicting design aspects with regards to the frame structure. Whereas further optimizations can be made for \ac{FDD} systems, the combination of sTTI and \ac{TDD} has limits.\looseness=-1

\textbf{FDD sTTI}: New features in Rel. 15 include slot and subslot configurations from Fig.~\ref{fig:FS1}. The \ac{DMRS} pattern is signaled in the UL \ac{DCI} and is used to reduce the DMRS overhead associated with the reduced TTI length. The DMRS symbol can be moved from the front to the end of a TTI or into the subsequent TTI using the different patterns. This allows sharing of one DMRS symbol among~TTIs.\looseness=-1

\textbf{TDD sTTI}: The original design of \textit{Frame structure type~3} in LTE did not cater for URLLC services. The minimal downlink-to-uplink switch-point periodicity is therefore \unit[5]{ms} in the uplink-downlink configurations $0, 1, 2$ and $6$. The configurations 3, 4 and 5 only support one downlink-to-uplink switching point with a periodicity of~\unit[10]{ms}~\cite{3gpp.36.211}.
This limits the minimal possible \ac{RTT} to two times the switching periodicity resulting in~\unit[10]{ms}~RTT. Future changes are unlikely due to backward compatibility issues. Therefore, the efforts to introduce URLLC in TDD-systems in the LTE standardization process was limited. Slot length sTTIs and reduced processing time of $n+3$ have been agreed in~\cite{3gpp.R1-1719247}. Although the URLLC target latency of \unit[1]{ms} remains unattainable for TDD LTE, the \unit[10]{ms} \ac{HRLLC} requirement can be met.\looseness=-1

\subsection{LTE Latency Calculation}
\label{sec:LTElatency}
\begin{figure}[b]
\vspace{-1mm}
\includegraphics[width=\linewidth]{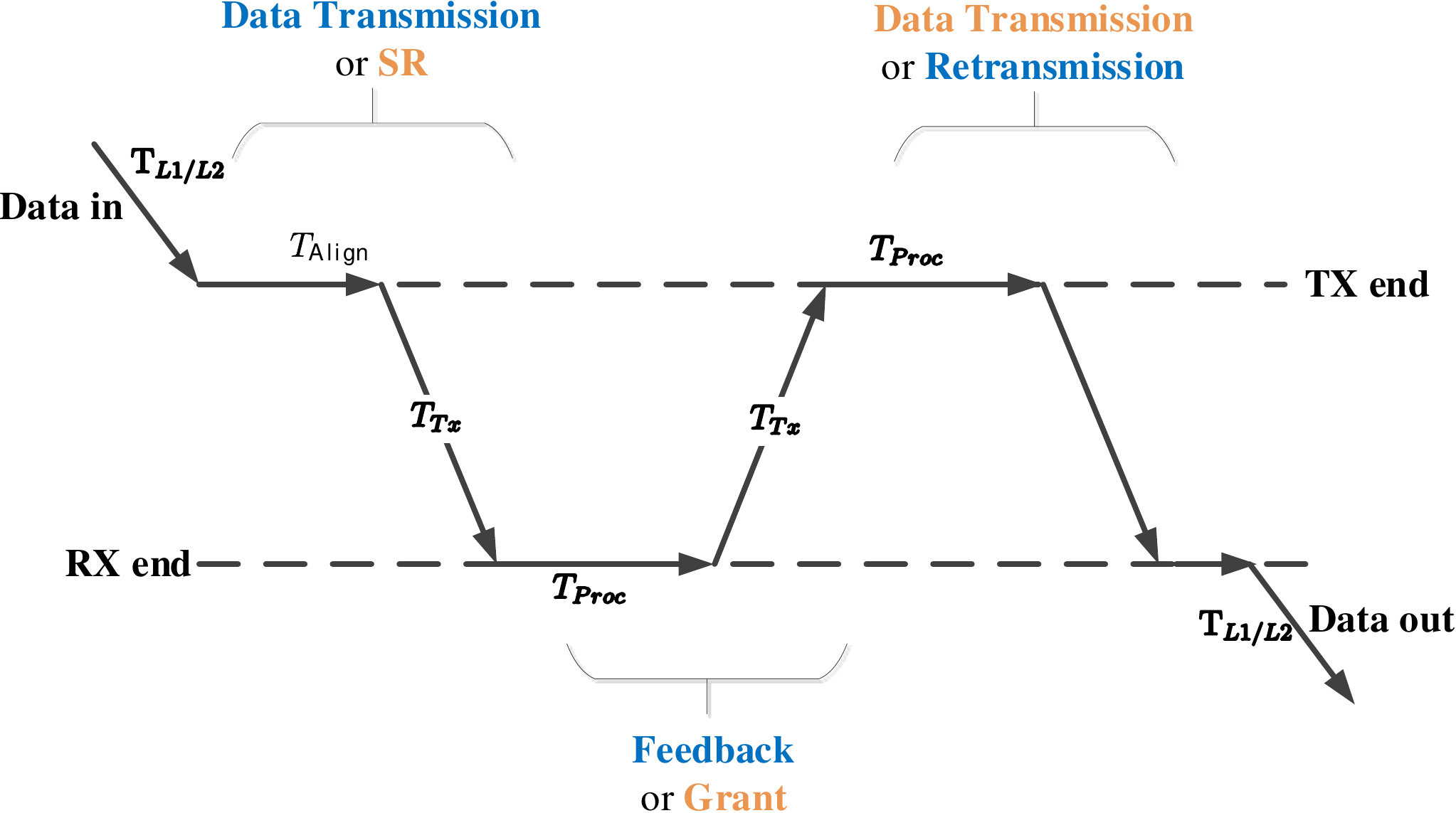}
  \caption{Illustration of latency components for DL (\textcolor[RGB]{10,89,178}{blue}) and UL (\textcolor[RGB]{227,130,65}{orange}) transmissions. The latency components are defined in Table~\ref{assumptions}.}
  \label{fig:latency}
\end{figure}

The ITU definition of user plane latency is the duration from L2/L3 ingress to L2/L3 egress~\cite{itu.m.2410}. Its timeline is depicted in Fig.~\ref{fig:latency}.
The definitions for the following delay analysis are shown in Table~\ref{assumptions}. It is assumed that the propagation time is significantly lower than one \ac{TTI}, and thus can be neglected. In case of \ac{HARQ}, HARQ-feedback and data retransmission can be repeated several times. This results in a total latency of:
\begin{equation}
\label{eq:T_total}
T_{\mathrm{Total}}= 2\cdot T_{\mathrm{L1/L2}}+T_{\mathrm{Align}}+ \sum{T_{\mathrm{Proc}}}+\sum{T_{\mathrm{Tx}}}.
\end{equation}
When using a repetition scheme without feedback, there is no $T_{\mathrm{Tx}}$ for the feedback, and the processing time is significantly shorter, reducing the sums for processing and transmission delay~\cite{3gpp.R1-1802882}.
\begin{table}[t]
\centering
\caption{E2E Latency Components related to Fig.~\ref{fig:latency}.}
\label{assumptions}
\begin{tabular}[width=\textwidth]{ll}
\hline
$T_{\mathrm{L1/L2}}$ & L1/L2 processing delay, \\ & for Rx and Tx at UE and eNB side respectively.\\
\rowcolor{gray!25}
$T_{\mathrm{Align}}$ & Alignment delay, the time required after \\
\rowcolor{gray!25}
& being ready to transmit and the transmission can start.\\
\rowcolor{gray!25}
& Worst-case latency is assumed (max. misalignment).\\

$T_{\mathrm{Proc}}$ & UE/eNB processing, time needed for preparing transmissions \\ & and decoding at the other side.  \\
\rowcolor{gray!25}
$T_{\mathrm{Tx}}$ & Transmission time.\\
\hline
\end{tabular}
\end{table}

\begin{table*}[t]

\centering
\caption{Latency results}
\label{tab:results}
\begin{subtable}{.49\textwidth}
\centering

\rowcolors{2}{gray!25}{white}
\begin{tabular}{llllll}
\hline
 &  & Rel. 14  & Rel. 15 & Rel. 15 & Rel. 15  \\
  &  & SF  & SF \& n+3 & slot &  subslot \\
 \hline\hline
\textbf{DL} & initial transmission 	& \yC 4  & \yC 4  & \yC 2 & \gC 0.7 \\
 & 1st retransmission 				& \rC 12 & \yC 10 & \yC 6 & \yC 2.0 \\
 & 2nd retransmission 				& \rC 20 & \rC 16 & \yC 10 & \yC 3.3 \\
 & 3rd retransmission 				& \rC 28 & \rC 22 & \rC 14 & \yC 4.7 \\
 \hline
\textbf{UL} & initial transmission 	& \rC 12 & \yC 10 & \yC 6 & \yC 2.0 \\
 & 1st retransmission 				& \rC 20 & \rC 16 & \yC 10 & \yC 3.3 \\
 & 2nd retransmission 				& \rC 28 & \rC 22 & \rC 14 & \yC 4.7 \\
 & 3rd retransmission 				& \rC 36 & \rC 28 & \rC 18 & \yC 6.0  \\ \hline
\end{tabular}
\label{tab:resultsHARQ}
\caption{with \ac{HARQ} retransmissions }
\end{subtable}
\begin{subtable}{.5\textwidth}
\centering

\rowcolors{2}{gray!25}{white}
\begin{tabular}{llllll}
\hline
 &  & Rel. 14 & Rel. 15 & Rel. 15 & Rel. 15 \\
 &  & SF  & SF \& n+3  & slot &  subslot \\
\hline\hline
\textbf{DL} & initial transmission 	& \yC 4 & \yC 4 & \yC 2 & \gC 0.7 \\
 & 1st repetition 					& \yC 5 & \yC 5 & \yC 2.5 & \gC 0.8 \\
 & 2nd repetition 					& \yC 6 & \yC 6 & \yC 3.0 & \gC 1.0 \\
 & 3rd repetition 					& \yC 7 & \yC 7 & \yC 3.5 & \yC 1.2 \\
 \hline
\textbf{UL} & initial transmission 	& \rC 12 & \yC 10 & \yC 6 & \yC 2.0 \\
 & 1st repetition 					& \rC 14 & \rC 12 & \yC 7 & \yC 2.3 \\
 & 2nd repetition 					& \rC 16 & \rC 14 & \yC 8 & \yC 2.7 \\
 & 3rd repetition 					& \rC 18 & \rC 16 & \yC 9 & \yC 3.0 \\ \hline
\end{tabular}
\label{tab:resultsHARQless}
\caption{HARQless repetition}
\end{subtable}

Calculated results for Downlink~(DL) and Uplink~(UL) for the LTE Rel. 14 SF (subframe) \unit[1]{ms} TTI as well as LTE Rel. 15 short processing time, slot and subslot configurations. The circles indicate the fulfillment of the \unit[10]{ms} \ac{HRLLC} (\yC) requirement and \unit[1]{ms} \ac{URLLC} (\gC) requirement respectively.
\end{table*}

\subsection{LTE Reliability Enhancements}

\begin{figure}[t]
\centering
\begin{subfigure}{.5\textwidth}
\centering
  \includegraphics[scale=0.3]{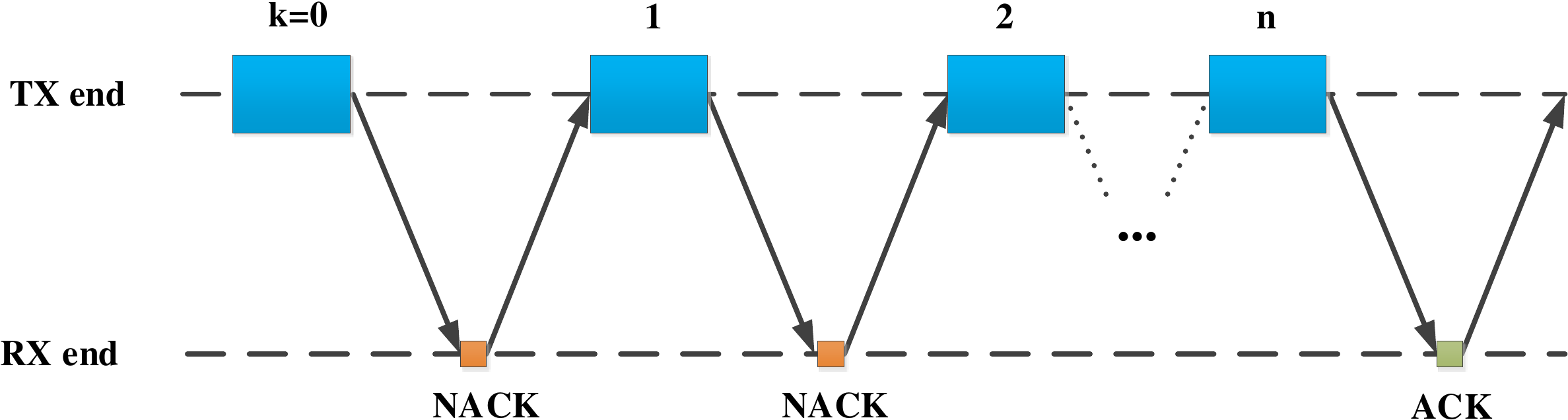}
  \caption{HARQ retransmission scheme with feedback and transmission of additional redundancy. After each iteration the receiver tries to decode and sends feedback (ACK/NACK) to the transmitter.}
  \label{fig:HARQscheme}
\end{subfigure}

\hfill

\begin{subfigure}{.5\textwidth}
\centering
  \includegraphics[scale=0.3]{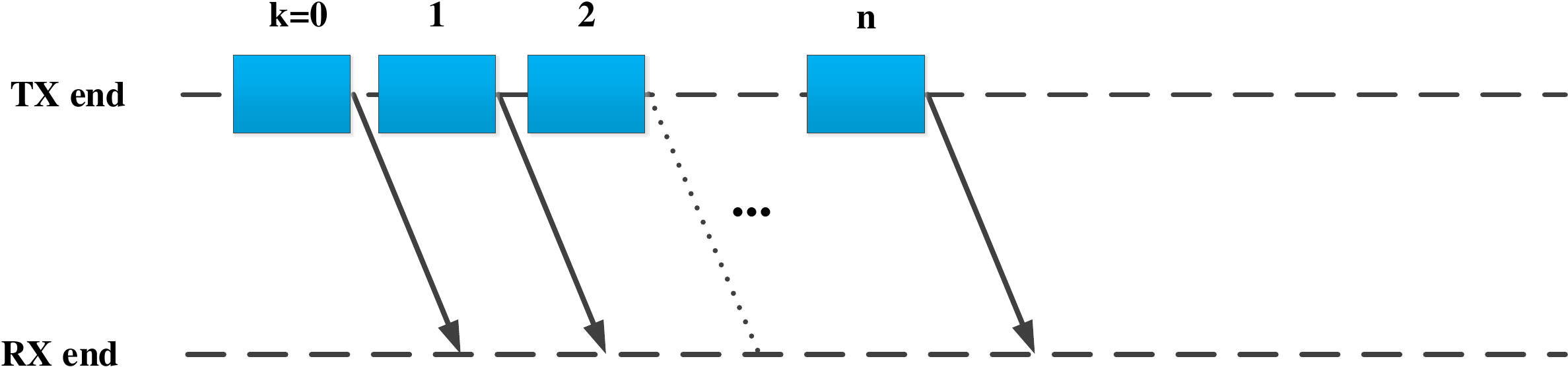}
  \caption{HARQless repetition scheme: a fixed number $k=n$ of repetitions is sent without waiting for feedback after each transmission.}
  \label{fig:HARQless}
\end{subfigure}

\caption{Retransmission schemes with $k$ as the number of retransmissions}
  \label{fig:HARQschemes}
\end{figure}

\textbf{LTE data channel} reliability in LTE is achieved by transmitting with a low code rate often split into separate transmissions. Additional redundancy is only transmitted when needed. This is also used for \ac{eMBB}-services to improve spectral efficiency.
The basic HARQ scheme is shown in Fig.~\ref{fig:HARQscheme}. In LTE, HARQ is configured with up to $k=3$ retransmissions~\cite{Dahlman}.
Alternatively, a set number of $k$ repetitions can be sent with an optional feedback at the end, Fig.~\ref{fig:HARQless}.
This scheme has less latency with the disadvantage of transmitting unnecessary redundancy versions compared to \ac{HARQ} with feedback.

\begin{figure}[b]
\centering
\tikzset{every picture/.style={line width=0.75pt}}

\begin{tikzpicture}[x=0.75pt,y=0.75pt,yscale=-1,xscale=1]

\draw   (243.5,87) .. controls (243.5,82.33) and (241.17,80) .. (236.5,80) -- (205.75,80) .. controls (199.08,80) and (195.75,77.67) .. (195.75,73) .. controls (195.75,77.67) and (192.42,80) .. (185.75,80)(188.75,80) -- (155,80) .. controls (150.33,80) and (148,82.33) .. (148,87) ;
\draw   (351.5,86) .. controls (351.5,81.33) and (349.17,79) .. (344.5,79) -- (313.75,79) .. controls (307.08,79) and (303.75,76.67) .. (303.75,72) .. controls (303.75,76.67) and (300.42,79) .. (293.75,79)(296.75,79) -- (263,79) .. controls (258.33,79) and (256,81.33) .. (256,86) ;
\draw    (257.25,144.7) -- (356.05,145.1) ;

\draw    (258.45,166.3) -- (357.25,166.7) ;

\draw  [dash pattern={on 0.84pt off 2.51pt}]  (159.5,111) -- (159.5,161) ;

\draw  [dash pattern={on 0.84pt off 2.51pt}]  (185,111) -- (185,161) ;

\draw  [dash pattern={on 0.84pt off 2.51pt}]  (228,111) -- (228,161) ;

\draw   (150.25,185) .. controls (150.25,189.67) and (152.58,192) .. (157.25,192) -- (185.25,192) .. controls (191.92,192) and (195.25,194.33) .. (195.25,199) .. controls (195.25,194.33) and (198.58,192) .. (205.25,192)(202.25,192) -- (233.25,192) .. controls (237.92,192) and (240.25,189.67) .. (240.25,185) ;
\draw    (249.5,207) .. controls (257.25,221) and (277.5,220) .. (294.5,193) ;
\draw [shift={(294.5,193)}, rotate = 483.71] [color={rgb, 255:red, 0; green, 0; blue, 0 }  ]   (0,0) .. controls (3.31,-0.3) and (6.95,-1.4) .. (10.93,-3.29)(0,0) .. controls (3.31,0.3) and (6.95,1.4) .. (10.93,3.29)   ;

\draw    (18, 57.33) rectangle (61.67, 224.99)   ;
\draw  [fill={rgb, 255:red, 207; green, 207; blue, 207 }  ,fill opacity=1 ] [rounded corners= 1.5] (21.08, 63.33) rectangle (58.58, 84.33)   ;
\draw  [fill={rgb, 255:red, 207; green, 207; blue, 207 }  ,fill opacity=1 ] [rounded corners= 1.5] (21.08, 90.33) rectangle (58.58, 111.33)   ;
\draw  [fill={rgb, 255:red, 207; green, 207; blue, 207 }  ,fill opacity=1 ] [rounded corners= 1.5] (21.08, 141.16) rectangle (58.58, 162.16)   ;
\draw    (75, 91) rectangle (133.5, 113)   ;
\draw    (39.83,100.83) -- (75,102) ;
\draw [shift={(75,102)}, rotate = 181.9] [color={rgb, 255:red, 0; green, 0; blue, 0 }  ]   (0,0) .. controls (3.31,-0.3) and (6.95,-1.4) .. (10.93,-3.29)(0,0) .. controls (3.31,0.3) and (6.95,1.4) .. (10.93,3.29)   ;

\draw    (133.5,102) .. controls (141.5,101) and (143.5,98) .. (149.5,95) ;
\draw [shift={(149.5,95)}, rotate = 510.17] [color={rgb, 255:red, 0; green, 0; blue, 0 }  ]   (0,0) .. controls (3.31,-0.3) and (6.95,-1.4) .. (10.93,-3.29)(0,0) .. controls (3.31,0.3) and (6.95,1.4) .. (10.93,3.29)   ;

\draw (158.8,95.2) node   {$d_{0}$};
\draw (187.6,95.6) node   {$d_{1}$};
\draw (231.6,94.4) node   {$d_{n}$};
\draw (210,92) node   {$...$};
\draw (269.4,96.2) node   {$s_{0}$};
\draw (298.2,95.6) node   {$s_{1}$};
\draw (342.2,95.4) node   {$s_{15}$};
\draw (320.6,93) node   {$...$};
\draw (197.5,64) node  [align=left] {control data};
\draw (305.5,63.5) node  [align=left] {scrambled CRC};
\draw (267.47,113.4) node   {$+$};
\draw (296.27,112.6) node   {$+$};
\draw (339.47,112.6) node   {$+$};
\draw (271.4,129.2) node   {$r_{0}$};
\draw (300.2,129.6) node   {$r_{1}$};
\draw (344.2,128.4) node   {$r_{15}$};
\draw (322.6,126) node   {$...$};
\draw (272,154.53) node [scale=0.5] [align=left] {mod\\ \ 2};
\draw (301,154.53) node [scale=0.5] [align=left] {mod\\ \ 2};
\draw (344.33,155.53) node [scale=0.5] [align=left] {mod\\ \ 2};
\draw (270.4,176.2) node   {$c_{0}$};
\draw (299.2,176.6) node   {$c_{1}$};
\draw (343.2,175.4) node   {$c_{15}$};
\draw (321.6,173) node   {$...$};
\draw (158.8,175.7) node   {$d_{0}$};
\draw (187.6,176.1) node   {$d_{1}$};
\draw (231.6,174.9) node   {$d_{n}$};
\draw (210,172.5) node   {$...$};
\draw (196.5,207) node  [align=left] {calculate CRC};
\draw (272.5,223.5) node [scale=0.8] [align=left] {check};
\draw (39.33,193.33) node [rotate=-269.84] [align=left] {PDCCH};
\draw (39,124) node  [align=left] {...};
\draw (105,102) node [scale=0.9] [align=left] {decoder};
\draw (41,73) node  [align=left] {{\footnotesize DCI}};

\end{tikzpicture}
\caption{DCI CRC blind decoding procedure in LTE~\cite{3gpp.36.212}.}
\label{fig:LTE_RNTI}
\end{figure}
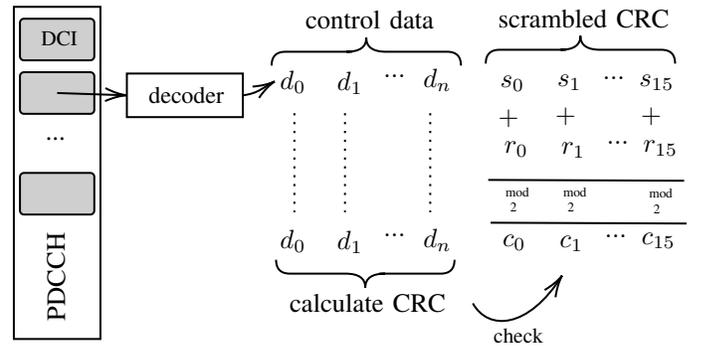
\textbf{LTE control channel}: The support of high reliability for LTE's control channel poses another design challenge. Control messages are sent as \acp{DCI} via the shared \ac{PDCCH}. This information is blind decoded and checked against a user specific \ac{RNTI}~\cite{3gpp.36.213}. Blind decoding may lead to false positive decoding of \acp{DCI} in the case of a coincidental but erroneous match.
A false positive can lead to successive errors, since the content of a control message is wrongly interpreted, also see~\cite{3gpp.R1-1719503}. The most serious error is buffer contamination of another transmission.
The procedure of blind decoding of a scrambled \ac{CRC} is depicted in Fig.~\ref{fig:LTE_RNTI}.

After decoding, the bits are split into the payload $d_0$ to $d_{n}$, and the scrambled CRC $s_0$ to $s_{15}$. After de-scrambling with the UE specific RNTI ($r_0$ to $r_{15}$), the result is compared with the calculated CRC of the received payload data.
If this does not match, either the decoding failed, or the payload is addressed to another UE with a different RNTI.
Blind decoding can lead to the false association of decoded or wrongly decoded \acp{DCI}~\cite{3gpp.R1-1802887} with a probability of

\begin{equation}
P_{\mathrm{FP}}=1-(1-2^{-16})^N.
\label{eq:FP}
\end{equation}

Here, $N$ is the number of blind decoding attempts and a uniform distribution is assumed for 16-bit \ac{CRC}.
With an assumption of $N=20$ blind decoding attempts, this results in a false positive rate of $P_{\mathrm{FP}}=3.05 \cdot 10^{-4}$. Note for HRLLC services, this is too high when targeting error rates below $10^{-4}$ for data transmissions.
There are two technical solutions proposed in 3GPP to reduce the false positive rate: firstly, increasing the \ac{CRC} length~\cite{3gpp.R1-1802180} and secondly \ac{DCI} duplication\cite{3gpp.R1-1802887}.

\begin{figure}[b]
	\includegraphics[width=\linewidth]{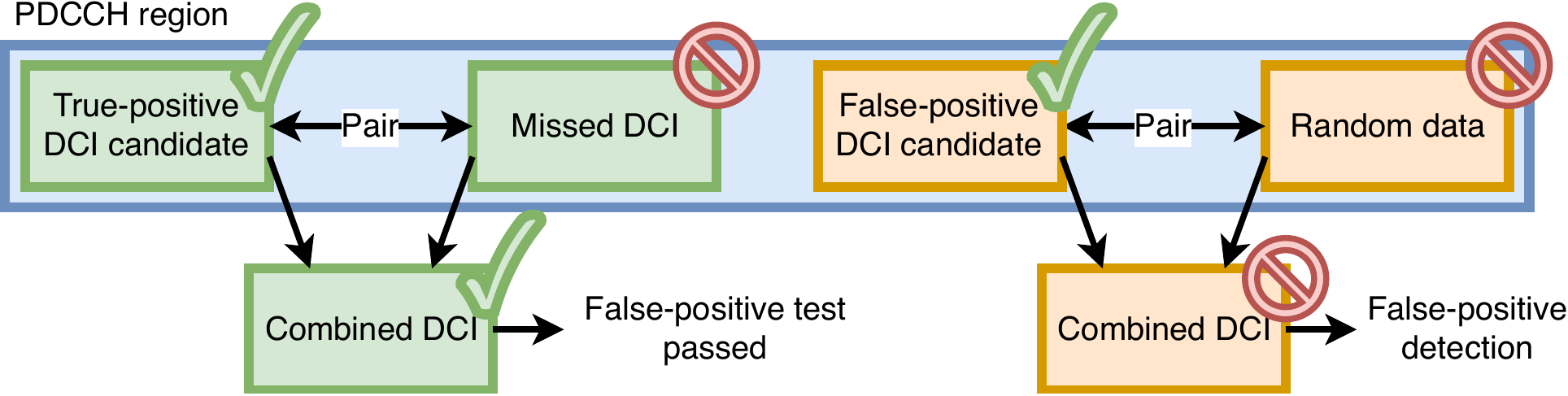}
	\caption{DCI Duplication}
	\label{fig:DCIdup}
\end{figure}

The DCI is currently limited to contain up to 8~\acp{CCE}, which limits the codeword size and thus lower bounds the code rate. Currently it is under discussion in 3GPP to double the number of~\acp{CCE} to 16 at low \acp{SNR}. However, this requires changes to the hashing function, see \cite{3gpp.R1-1801941}. Alternatively, two duplicate \acp{DCI} can be sent and simultaneously used to enable operation at low SNRs and improve the rejection of false positives as shown in Fig. \ref{fig:DCIdup}:
\begin{itemize}
\item On the left, a UE receives two DCIs but misses one. Here the two are combined and the resulting combined DCI is valid.
\item On the right, a random DCI is falsely decoded and passes the CRC check. Here the combination is different and not valid.
\end{itemize}

\section{Performance Evaluation And Analysis}
Next, we analyze the available LTE configurations with respect to the \ac{URLLC} or the less stringent HRLLC requirements. Eq.~\ref{eq:T_total} in the previous section describes the total over-the-air latency. For LTE, the latency analysis has to differentiate between {DL} and {UL}, since initially a scheduling request has to be sent for acquiring {UL} resources. The latency in downlink~(DL) direction for a transmission using HARQ~(HA) with $k$ retransmissions can be obtained as
\begin{equation}
T_{\mathrm{DL},\mathrm{HA}}=T_{c}+2\cdot k\cdot T_{\mathrm{Proc}}+(1+2\cdot k)\cdot T_{\mathrm{Tx}}.
\end{equation}
Here, the delay caused by the higher layers and alignment remains constant with $T_{c}=2\cdot T_{\mathrm{L1/L2}}+T_{\mathrm{Align}}$. For the uplink~(UL) direction, there is an additional \ac{RTT} due to the scheduling request~(SR) and following uplink grant. This leads to $k_{\mathrm{UL}}=k+1$, and thus the UL delay including HARQ results to
\begin{equation}
T_{\mathrm{UL},\mathrm{HA}}=T_{\mathrm{c}}+2\cdot k_{\mathrm{UL}}\cdot T_{\mathrm{Proc}}+(1+2\cdot k_{\mathrm{UL}})\cdot T_{\mathrm{Tx}}.
\end{equation}
For HARQ-less~(HL) repetitions, the absence of feedback reduces the latency. Thus for $k$ repetitions, the latency in the \ac{DL} can be defined as
\begin{equation}
T_{\mathrm{DL},\mathrm{HL}}=T_{\mathrm{c}}+(1+k)\cdot T_{\mathrm{Tx}},
\end{equation}
and for UL it will be
\begin{equation}
T_{\mathrm{UL},\mathrm{HL}}=T_{\mathrm{c}}+2\cdot T_{\mathrm{Proc}}+(1+2\cdot k_{\mathrm{UL}})\cdot T_{\mathrm{Tx}}.
\end{equation}
Here, $k_{\mathrm{UL}}=k+1$ again results from the scheduling request and grant. For comparison, we calculate the \ac{E2E} delays to obtain quantitative results. For this, the following assumptions are made:
\begin{itemize}
\item $T_{\mathrm{L1/L2}}=1 \, \textrm{TTI}$,
\item $T_{\mathrm{Align}}=1 \, \textrm{TTI}$, as a reception arriving just after a TTI starts needs to be delayed for one TTI,
\item $T_{\mathrm{Proc}}=3 \, \textrm{TTI}s$, unless using the reduced processing time feature for which $T_{\mathrm{Proc}}=2 \, \textrm{TTI}$,
\item $T_{\mathrm{Tx}}=1 \, \textrm{TTI}$, transmissions spanning 1 TTI.
\end{itemize}
Using the normal cyclic prefix~(CP), each LTE subframe contains 14~OFDM symbols with a duration of~\unit[1]{ms}. This results in the following TTI lengths for subframe (sf), slot and subslot:

\begin{itemize}
\item $T_{\textrm{sf TTI}}$ \tabto{1.57cm} $=1 \, \textrm{TTI}$ \tabto{3.3cm} $= \unit[1]{ms},$
\item $T_{\textrm{slot TTI}}$ \tabto{1.57cm} $= T_{\textrm{sf TTI}}/2$ \tabto{3.3cm} $= \unit[0.5]{ms},$
\item $T_{\textrm{subslot TTI}}$ \tabto{1.57cm} $= T_{\textrm{sf TTI}}/6$ \tabto{3.3cm} $= \unit[0.17]{ms}.$
\end{itemize}

\begin{table}[ht]
\centering
\caption{Simulation assumptions for DCI duplication}
\label{tab:DCIdupSim}
\rowcolors{1}{gray!25}{white}
\begin{tabular}{ll}
\hline
Channel Model     & Rayleigh fading (ideal channel estimation)                                                                        \\
DCI payload       & 45 bits                                                                                                           \\
CRC size          & 16 bits                                                                                                           \\
DCI blind decodes & 20                                                                                                                \\
Channel Code      & TBCC AL 1-8                                                                                                       \\
Decoder           & Viterbi                                                                                                           \\
Chase Combining   & Bitwise: LLRs are combined Bitwise
                                          \\
\rowcolor{gray!25}
& Symbolwise: combining of QAM-Symbols
                                          \\
\hline
\end{tabular}
\end{table}

Table~\ref{tab:results} lists the calculated latencies for 3GPP LTE~Rel.~14 and 15 for subframe~(SF), slot, and subslot configurations.\\
It can be seen, that with HARQ, only the~\unit[10]{ms} HRLLC requirement is within reach when using the LTE~Rel.~15 subslot configuration.
HARQ-less repetition improves performance and brings \unit[1]{ms} URLLC into reach for \ac{DL} transmission with the LTE Rel.~15 subslot configuration. In the UL, the delay caused by the \ac{SR} is too high. The less stringent \unit[10]{ms} HRLLC requirement makes UL possible for Rel.~15 slot and subslot configurations. In the DL, even LTE~Rel.~14 subframe configuration fulfills the delay requirements.
In LTE, UL is handled by using \ac{SPS} with pre-allocated resources thereby removing the additional delay caused by SR and grant time.

Next, we evaluate the reliability of the control channel when using the proposed DCI duplication mechanism. The performance of DCI duplication is numerically evaluated using LTE link-level simulations. Details are given in Table~\ref{tab:DCIdupSim}. For this, a 16-bit \ac{CRC} is added to a generated 45-bit DCI, which are then sent over a Rayleigh fading channel. For combining the two duplicate DCIs, we compare two schemes:
\begin{itemize}
\item Bitwise: \acp{LLR} are combined bitwise before decoding,
\item Symbolwise: received \ac{QAM}~symbols are combined before demodulation.
\end{itemize}
The code rate is varied by changing the \ac{AL} which defines how many \acp{CCE} are used for transmission. Thus, a higher AL effectively decreases the code rate.
For the link-level simulations, a fixed pairing of duplicate DCIs is assumed and chase combining is only performed if one of the two DCIs is correctly decoded. Thus, a DCI is missed either if both initial decodes fail, or if the combination of the two decodes does not result in the correct DCI. This is also referred to as \textit{miss probability}.

\begin{figure}
    \centering
    \begin{subfigure}[b]{0.47\textwidth}
    \includegraphics[width=\textwidth]{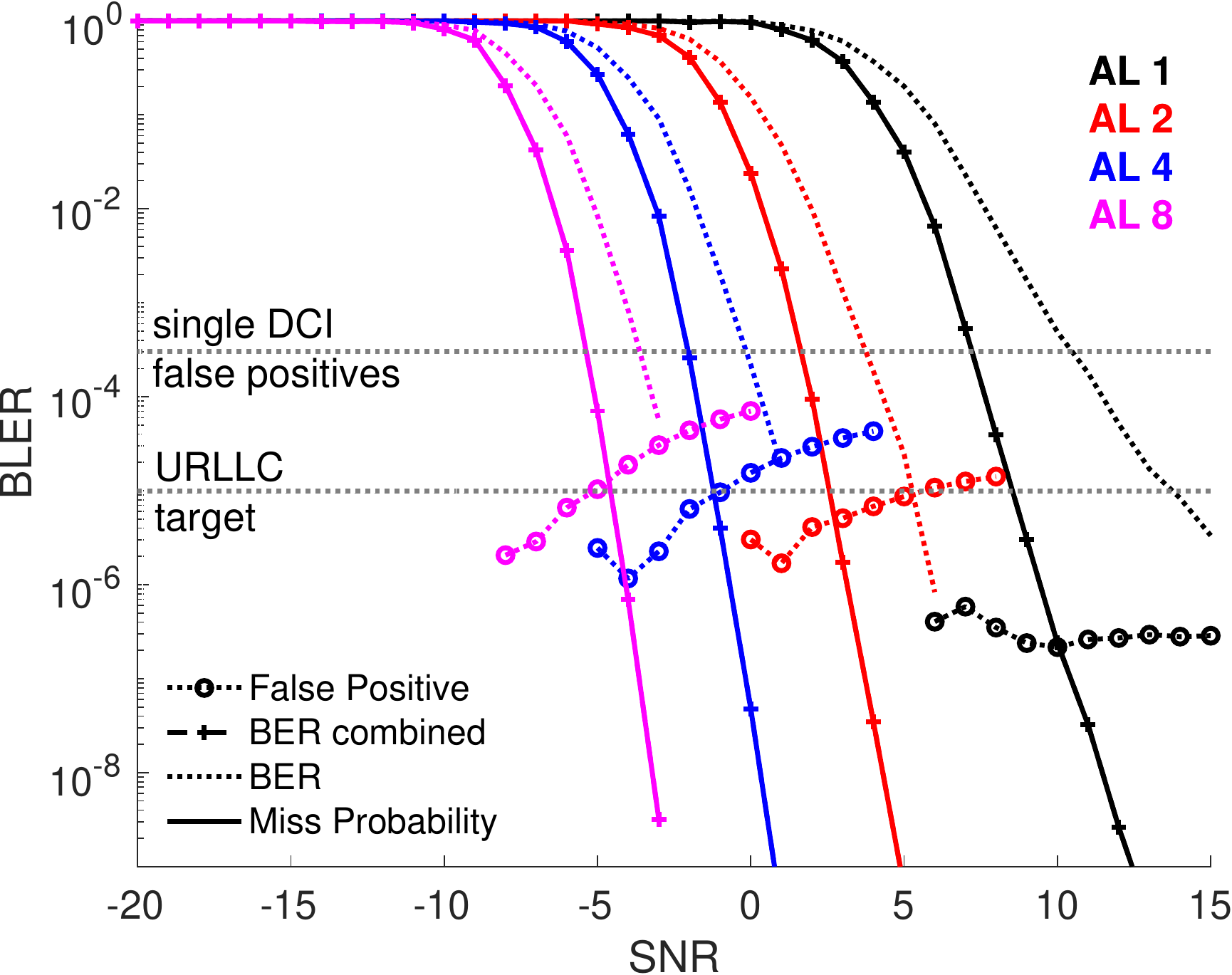}
    \caption{Bitwise LLR combining}
    \label{fig:DCIdupPlotBit}
    \end{subfigure}

    \begin{subfigure}[b]{0.47\textwidth}
    \includegraphics[width=\textwidth]{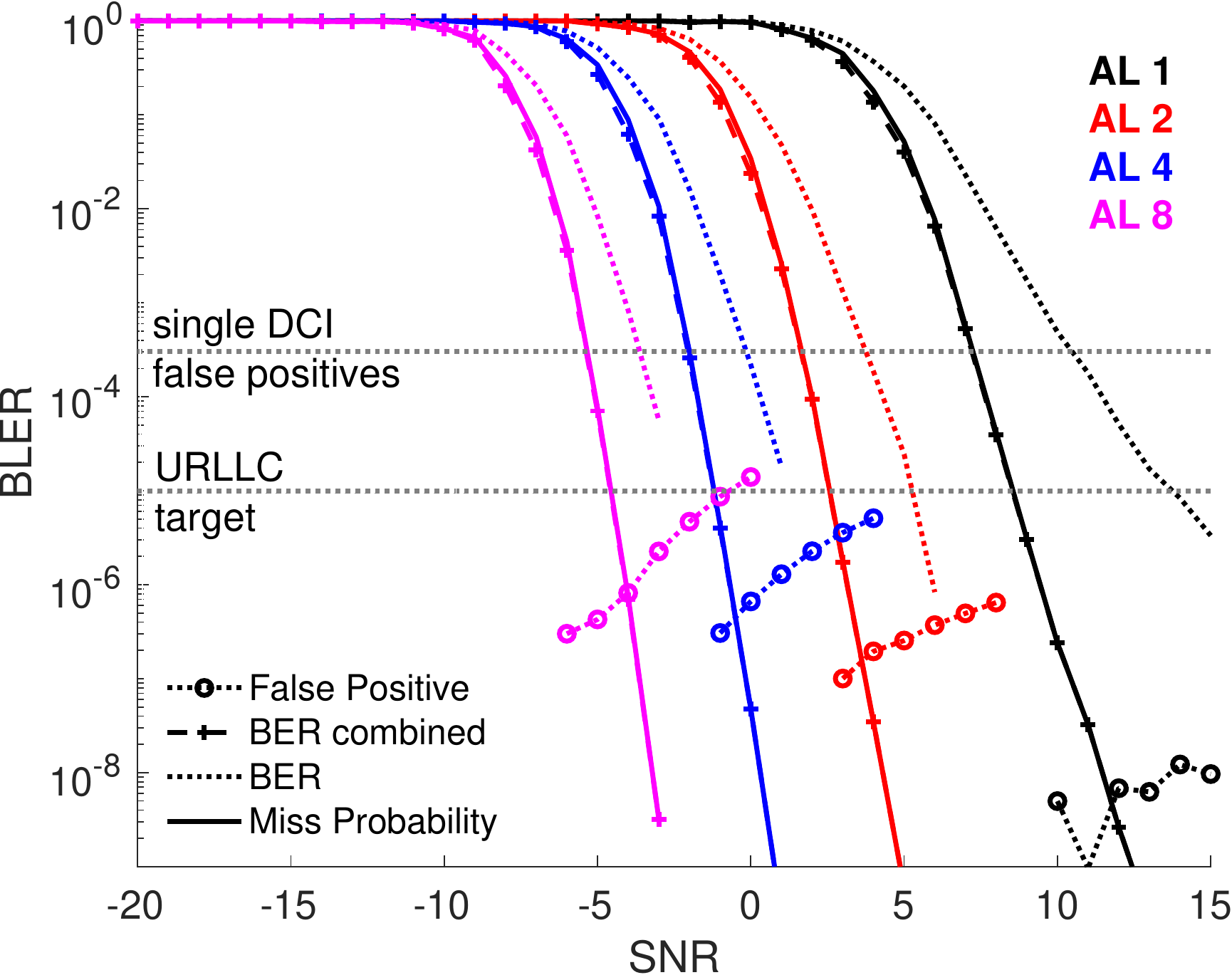}
    \caption{QAM symbolwise combining}
    \label{fig:DCIdupPlotSym}
    \end{subfigure}
    \caption{Numerical results on DCI duplication comparing different combining methods.}\label{fig:DCIdupPlots}
\end{figure}

The results are shown in Fig.~\ref{fig:DCIdupPlots}. As a reference, the single DCI false positive probability is shown by the gray dotted line. This is analytically calculated from Eq.~\ref{eq:FP} with 20~blind decoding attempts, which is also used in the link-level simulations. In addition to the false positives, the \ac{BLER} of a single DCI is compared to the \ac{BLER} of combined DCIs in Fig.~\ref{fig:DCIdupPlots}(a) and (b). With both schemes it can be seen, that the assumption of only performing the chase combining upon the detection of at least one DCI, does not significantly affect the performance. The combined BLER and the miss probability, also considering false rejection, are the same at the targeted error rates over all \acp{AL}. Finally, the QAM-symbolwise combining of DCIs in Fig.~\ref{fig:DCIdupPlots}(b) suppresses false positives significantly better.

\pagebreak

\section{From LTE to New Radio (NR)}
\ac{LTE}'s basic frame structure design was fixed with \ac{LTE} Rel. 8. Although the control channel can vary from 1 to 3~OFDM symbols, based on the \ac{CFI}, its size and position within a radio frame are fixed. Furthermore, the control channel spans over the whole frequency domain, forcing each \ac{UE} to perform blind decoding over the whole frequency band, e.g. the maximum is over one component carrier, which is 20~MHz bandwidth. Future LTE releases have to stay backwards compatible to the existing radio frame structure. This static design limits implementation of new URLLC or HRLLC services, especially when multiplexing services with different service requirements, such as eMBB and URLLC in the same frequency band. Especially when it comes to TDD, \ac{LTE}'s frame structure is quite limited, with only 8 different TDD modes. With this fixed number of time slots in up- or downlink, for a closed-loop service, a constant delay in the range of milliseconds is added to each transmission.

The goal of \ac{NR} is to overcome the design limits of LTE by defining a "forward-compatible" frame structure. The idea is to define the frame structure in such a way, that new services can easily be added in the future. The key ingredients to support \ac{URLLC} service in \ac{NR} are mini-slots, a self-contained frame structure, and grant-free radio access concepts \cite{takeda2017latency}. Similar to \ac{sTTI}, \ac{NR} supports a short subslot format, called mini-slots or non-slot based scheduling. The self-contained frame structure allows a UE to only decode a very short control channel organized in control-resource sets~(CORESETs) with a UE specific search space prior to decoding the data channel. This reduces the processing time and allows fast feedback to the transmitter based on the decoding outcome. Furthermore, coding rates of down to $1/12$ are expected for \ac{URLLC} channel coding \cite{3gpp.R1-163757}. Due to the expected sporadic nature of \ac{URLLC} traffic, a new multiplexing concept based on pre-emption has been introduced in \ac{NR} \cite{3gpp.R1-1716941}. This allows puncturing of \ac{eMBB} transmissions in case that unexpected \ac{URLLC} traffic arrives at least for \ac{DL}. Since the \ac{eMBB} transmission is degraded by this mechanism, a new report, so-called \ac{PI}, is introduced to indicate punctured positions afterwards. The same concept is currently discussed for the \ac{UL} \cite{3gpp.R1-1801566}.

Although, \ac{HARQ} timing is designed flexibly in \ac{NR} \cite{3gpp.R1-1716941}, the main issue of \ac{HARQ} \ac{RTT} is still an open topic. To cope with this limitation, the authors of \cite{ldpc_subcodes} have proposed an early \ac{HARQ} feedback technique, which enables to start processing during reception. Hence, the receiver can provide the feedback at an earlier stage. This early feedback is enabled by exploiting substructures of the channel code. As shown in \cite{ldpc_subcodes}, subcode-based early HARQ achieves a reliability comparable to regular HARQ while decreasing the \ac{HARQ} \ac{RTT}.

\section{Conclusion}
5G is envisioned to support three broad categories of services: eMBB, URLLC, and mMTC. URLLC services refer to future applications which require secure data communications from one end to another, while fulfilling ultra-high reliability and low latency. These have been addressed in 3GPP Rel.~15 for LTE by two work items~(WIs), as well as considered in the basic design of NR.
This paper gives an overview of URLLC requirements and describes technical innovations and standardization efforts in 3GPP LTE.
Latency reduction techniques with reduced processing time and improved frame structures, shortened TTI, are shown.
Furthermore, a detailed analysis of the resulting latencies, which are feasible with LTE Rel. 15 are given. Especially, the reliability limits of LTE's control channel are highlighted and solutions are presented. Our numerical results show that QAM-symbol combining of duplicate DCIs improves control channel robustness achieving URLLC targets.
Our presented solution can also be adopted for improving the robustness of the control channel in „forward-compatible“ 5G \ac{NR} systems.

    \bibliographystyle{IEEEtran}
	\bibliography{3gpp,references,RAN1_92,RAN1_90b}{}

\begin{thebibliography}{10}
\providecommand{\url}[1]{#1}
\csname url@samestyle\endcsname
\providecommand{\newblock}{\relax}
\providecommand{\bibinfo}[2]{#2}
\providecommand{\BIBentrySTDinterwordspacing}{\spaceskip=0pt\relax}
\providecommand{\BIBentryALTinterwordstretchfactor}{4}
\providecommand{\BIBentryALTinterwordspacing}{\spaceskip=\fontdimen2\font plus
\BIBentryALTinterwordstretchfactor\fontdimen3\font minus
  \fontdimen4\font\relax}
\providecommand{\BIBforeignlanguage}[2]{{%
\expandafter\ifx\csname l@#1\endcsname\relax
\typeout{** WARNING: IEEEtran.bst: No hyphenation pattern has been}%
\typeout{** loaded for the language `#1'. Using the pattern for}%
\typeout{** the default language instead.}%
\else
\language=\csname l@#1\endcsname
\fi
#2}}
\providecommand{\BIBdecl}{\relax}
\BIBdecl

\bibitem{path25G}
{K. Mallinson}, ``The path to 5{G}: as much evolution as revolution,'' Tech.
  Rep., May 2016.

\bibitem{itu.m.2410}
{ITU-R}, ``Minimum requirements related to technical performance for {IMT}-2020
  radio interface(s),'' {ITU - International Telecommunication Union}, Tech.
  Rep., {N}ov. 2017.

\bibitem{20155GWhitePaperNGMN}
{NGMN Alliance}, ``{NGMN 5G} white paper,'' Feb. 2015.

\bibitem{7980747}
C.-P. Li, J.~Jiang, W.~Chen, T.~Ji, and J.~Smee, ``{5G} ultra-reliable and
  low-latency systems design,'' in \emph{2017 European Conference on Networks
  and Communications (EuCNC)}, June 2017, pp. 1--5.

\bibitem{3GP15Timeline}
{3GPP}, ``{SP-150149: “5G”} timeline in {3GPP},'' 2015.

\bibitem{3gpp.RP-171489}
3GPP, ``Work item on ultra reliable low latency communication for {LTE},'' {3rd
  Generation Partnership Project (3GPP)}, TDoc RP-171489, {June} 2017.

\bibitem{3gpp.36.881}
------, ``Study on latency reduction techniques for {LTE},'' {3rd Generation
  Partnership Project (3GPP)}, Technical Report (TR) 36.881, {July} 2016,
  version 14.0.0.

\bibitem{3gpp.RP-161299}
------, ``Work item on shortened {TTI} and processing time for {LTE},'' {3rd
  Generation Partnership Project (3GPP)}, TDoc RP-161299, {June} 2017.

\bibitem{3gpp.38.913}
------, ``Study on scenarios and requirements for next generation access
  technologies,'' {3rd Generation Partnership Project (3GPP)}, Technical Report
  (TR) 38.913, {March} 2017, version 14.2.0.

\bibitem{2015AdvancedSDRWirth}
T.~Wirth, M.~Mehlhose, J.~Pilz, R.~Lindstedt, D.~Wieruch, B.~Holfeld, and
  T.~Haustein, ``An advanced hardware platform to verify {5G} wireless
  communication concepts,'' in \emph{Proc. of IEEE VTC-Spring}, May 2015.

\bibitem{2016TacintPilz}
J.~Pilz, M.~Mehlhose, T.~Wirth, D.~Wieruch, B.~Holfeld, and T.~Haustein, ``A
  tactile internet demonstration: 1ms ultra low delay for wireless
  communications towards {5G},'' in \emph{2016 IEEE Conference on Computer
  Communications Workshops (INFOCOM)}, April 2016.

\bibitem{3gpp.36.211}
3GPP, ``Evolved universal terrestrial radio access ({E-UTRA}); physical
  channels and modulation,'' {3rd Generation Partnership Project (3GPP)},
  Technical Specification (TS) 36.211, {March} 2017, version 14.2.0.

\bibitem{3gpp.R1-1719247}
Ericsson, ``{RAN1} decisions for {WI} shortened {TTI} and processing time for
  {LTE17},'' {3rd Generation Partnership Project (3GPP)}, TDoc R1-1719247,
  {Nov} 2017.

\bibitem{3gpp.R1-1802882}
------, ``Latency for {URLLC},'' {3rd Generation Partnership Project (3GPP)},
  TDoc R1-1802882, {Feb} 2018.

\bibitem{Dahlman}
S.~P. Erik~Dahlman and J.~Sköld, ``In {4G} {LTE-A}dvanced {P}ro and the road
  to {5G} ({T}hird {E}dition),'' 2016.

\bibitem{3gpp.36.212}
3GPP, ``Evolved universal terrestrial radio access ({E-UTRA}); multiplexing and
  channel coding,'' {3rd Generation Partnership Project (3GPP)}, Technical
  Specification (TS) 36.212, {March} 2017, version 14.2.0.

\bibitem{3gpp.36.213}
------, ``Evolved universal terrestrial radio access ({E-UTRA}); physical layer
  procedures,'' {3rd Generation Partnership Project (3GPP)}, Technical
  Specification (TS) 36.213, {March} 2017, version 14.2.0.

\bibitem{3gpp.R1-1719503}
------, ``Design impact on reliability for {LTE URLLC},'' {3rd Generation
  Partnership Project (3GPP)}, TDoc R1-1719503, {Nov} 2017.

\bibitem{3gpp.R1-1802887}
{Fraunhofer HHI}, ``{DL} control techniques for {LTE URLLC},'' {3rd Generation
  Partnership Project (3GPP)}, TDoc R1-1802887, {Feb} 2018.

\bibitem{3gpp.R1-1802180}
L.~Electronics, ``Candidate techniques for {DL} control for {LTE} {URLLC},''
  {3rd Generation Partnership Project (3GPP)}, TDoc R1-1802180, {Feb} 2018.

\bibitem{3gpp.R1-1801941}
Samsung, ``Discussion on {DL} control related techniques for {URLLC},'' {3rd
  Generation Partnership Project (3GPP)}, TDoc R1-1801941, {Feb} 2018.

\bibitem{takeda2017latency}
K.~Takeda, L.~H. Wang, and S.~Nagata, ``Latency reduction toward 5{G},'' in
  \emph{{2017 IEEE Wireless Communication}}, June 2017.

\bibitem{3gpp.R1-163757}
Nokia, ``{WF} on channel coding evaluation for {5G} new radio,'' {3rd
  Generation Partnership Project (3GPP)}, TDoc R1-163757, {April} 2016.

\bibitem{3gpp.R1-1716941}
ETSI, ``Report of {RAN}1\#90 meeting,'' {3rd Generation Partnership Project
  (3GPP)}, TDoc R1-1716941, {Oct} 2017.

\bibitem{3gpp.R1-1801566}
Ericsson, ``On pre-emption in uplink,'' {3rd Generation Partnership Project
  (3GPP)}, TDoc R1-1801566, {Feb} 2018.

\bibitem{ldpc_subcodes}
B.~G\"{o}ktepe, S.~F\"{a}hse, L.~Thiele, T.~Schierl, and C.~Hellge,
  ``Subcode-based early {HARQ} for {5G},'' in \emph{{2018 IEEE International
  Conference on Communications Workshops (ICC)}}, May 2018.

\end{thebibliography}

\end{document}